\documentclass[namedreferences]{kluwer}
\usepackage{graphicx}

\begin{document}
\begin{article}

\runningtitle{Dynamics of electron beams in the inhomogeneous solar corona
plasma}

\runningauthor{Kontar, E.P.}

\begin{opening}

\title{Dynamics of electron beams in the inhomogeneous solar corona plasma}

\subtitle{}
\author{Eduard P. \surname{Kontar}}
\institute{Institute of Theoretical Astrophysics, P.B. 1029,
Blindern, 0315 Oslo, Norway}
\date{\today}
\begin{abstract}
Dynamics of an spatially limited electron beam in the
inhomogeneous solar corona plasma is considered in the framework
of weak turbulence theory when the temperature of the beam
significantly exceeds that of surrounding plasma. The numerical
solution of kinetic equations manifests that generally the beam
accompanied by Langmuir waves propagates as a beam-plasma
structure with a decreasing velocity. Unlike the uniform plasma
case the structure propagates with the energy losses in the form
of Langmuir waves. The results obtained are compared with the
results of observations of type III bursts. It is shown that the
deceleration of type III sources can be explained by the corona
inhomogeneity. The frequency drift rates of the type III sources
are found in a good agreement with the numerical results of beam
dynamics.

\end{abstract}

\end{opening}

\section{Introduction}

Type III bursts seem to be the most intensively studied type of
solar sporadic radio emission. The theory of these bursts involves
different aspects of solar physics, plasma physics, and nonlinear
physics. Current understanding of the problem was summarized in
the following reviews
\cite{Goldman83,Dulk85,Melrose90,Muschietti90}. Thus, fast
electrons propagating along open magnetic field lines become a
source of Langmuir waves. In their turn, Langmuir waves are partly
transformed into observable radio emission via nonlinear plasma
processes. Propagation of electron beams accelerated during the
solar flares is considered to be one of the key problems in the
theory of type III solar radio bursts \cite{Melrose90}. The
difficulty of the problem originates from the fact that electrons
propagating in a collisionless plasma generate Langmuir waves,
which influence on the electron propagation. For most type III
events, the level of Langmuir waves generated by the beam is quite
low \cite{Grognard85} and the beam dynamics is considered in the
framework of weak turbulence theory \cite{Vedenov62,Drummond62}.
Even in the simplest, one-dimensional case the system of kinetic
equations cannot be solved without simplifying assumptions
\cite{Ryutov70}. Therefore, there are many works
treating this problem analytically \cite{Ryutov70, Zaitsev72,
Grognard75, Takakura76, Melnik95}  as well as numerically
\cite{Smith77, Takakura76, Grognard80, Takakura82, Grognard85,
Melnik99}.

The presence of coronal plasma inhomogeneity is the important
feature in the description of type III solar radio bursts. The
electron beam propagating in the plasma with decreasing density
(i.e. decreasing local plasma frequency) can emit at the plasma
(fundamental) and double plasma (harmonic) frequencies. The
relation between the drift rate $D_f$ measured at the local plasma
frequency $f=\omega _{pe}/2\pi$ and the radial source velocity
$V_r$ is given by
\begin{equation}\label{drift}
  D_f =\frac{\mbox{d}f}{\mbox{d}t}= \frac fL V_r,\;\; \mbox{where}\;\; L\equiv\left(
  \frac{1}{\omega _{pe}}\frac{\mbox{d} \omega_{pe}}{\mbox{d} r}\right)^{-1}=\left(
  \frac 1{2n(r)}\frac{\mbox{d}n}{\mbox{d}r}\right)^{-1}
\end{equation}
where $r$ is the radial distance from the center of the Sun,
$\omega _{pe}=\sqrt{4\pi e^2n(r)/m}$ is the local plasma
frequency, $n(r)$ is the plasma electron density. The presence of
the local plasma frequency gradient leads to two physical effects in
the kinetics of Langmuir waves:

i) The Langmuir wave propagating in the inhomogeneous plasma
experiences a shift of wavenumber $\Delta k(r)$, due to the
variation of the local refractive index.

ii) The time of beam-plasma interaction $\tau (r)$ (quasilinear
time) depends on the local density and therefore the resonance
condition for the plasmons may itself change during the course of
beam propagation.

The first effect is considered to have main impact on Langmuir
wave kinetics and effect ii) can be neglected in comparison with
effect i) \cite{Coste75}.

The influence of plasma inhomogeneity is usually considered to be
a suppressing process for quasilinear relaxation
\cite{Ryutov69,Breizman69,Krasovskii78}. It was shown that the
inhomogeneity of plasma can significantly change the rate  of
quasilinear relaxation for low-density electron beams, when the
length of quasilinear relaxation is comparable with the
characteristic length of plasma inhomogeneity. Langmuir waves
generated by the beam move out of the resonance velocities that
decrease wave generation and prevent plateau formation at the
electron distribution function.

If we assume that the electron beam generating type III bursts has
low density then the beam can propagate in the state close to
marginal stability giving rise to Langmuir waves due to plasma
density fluctuations \cite{Robinson92b,Robinson93}. On the
contrary there are some evidences \cite{Thejappa98} that electron
beams associated with the bursts are so dense that they can
generate Langmuir waves in strong turbulence limit. Therefore, it
is natural to take an intermediate beam density for consideration.
Thus, the electron beam is dense enough to form a quasilinear
steady state but the level of plasma turbulence is within the weak
turbulence limit. For such electron beams the quasilinear length
is much smaller than that of plasma inhomogeneity and the effects
connected with the plasma inhomogeneity are usually not taken into
account. However, at the starting frequencies of type III bursts
$f>200$MHz the drift rate is high (and consequently the plasma
density gradient) and the exclusion of the plasma gradient cannot be
justified. Therefore, to achieve the quantitative understanding in
theory of type III bursts one has to include the influence of
plasma density gradient.

 For a theoretical descriptions of experiments in laboratory
 plasma \cite{Ryutov69, Breizman69, Krasovskii78} and some astrophysical
 applications \cite{McClements89} it is necessary to consider
 the stationary boundary-value problems involving the injection
 of a beam into a plasma half-space. However, for the interpretation and
 theoretical description of type III bursts the initial value
 problem seems to be more natural.

 In this paper we investigate the dynamics of spatially limited
 fast electron beam in the inhomogeneous solar
 corona plasma with monotonically decreasing density.
 In our calculations we use quite realistic isothermal solar corona density
 model \cite{Mann99}. The consideration of electron propagation is
 based on the system of quasilinear equation of weak turbulence
 theory where the influence of plasma density gradient is included.
For the first time the dynamics of the electron cloud is
investigated at large distances when the influence of plasma
inhomogeneity is taken into account.
 We performed numerical computations of the beam dynamics for a wide range of initial
  beam parameters: initial velocity of the beam $v_0$ from $0.3c$ to $0.75c$
 and initial beam density $n'$ from $1$~cm$^{-3}$ to $500$~cm$^{-3}$.
 These parameters allows us to obtain the source velocities in the
 wide range from $0.15c$ to $0.75c$ that covers the majority of
 observational data. The results obtained are compared
 with the results of observations.

\section{Quasilinear equations for inhomogeneous plasma}

Let us consider the propagation of the electron beam cloud when
the energy density of excited Langmuir waves is much less than
that of surrounding plasma
\begin{equation}
W/nT \ll (k\lambda _{D})^2,
 \label{eq1}
\end{equation}
where $W$ is the energy density of Langmuir waves, $T$ is the
temperature of the  surrounding plasma, $k$ is the wave number,
and $\lambda _{D}$ is the electron Debye length.  Our analysis uses
one-dimensional kinetic equations following
Ryutov and Sagdeev (\citeyear{Ryutov70}), Zaitsev, Mityakov,
and Rapoport (\citeyear{Zaitsev72}), and Grognard (\citeyear{Grognard85}).
Firstly, the one-dimensional
character of beam propagation is supported by numerical solution
of 3D equations \cite{Churaev80}. Secondly, in the application to
III type solar bursts  electrons propagate along the magnetic field
and $H^2/8\pi \gg nmv_0^2/2$ \cite{Dulk85} that ensures
one-dimensional character of electron propagation.

In this case, as was shown in \cite{Vedenov62,Drummond62} one
can use equations of quasilinear theory
\begin{equation}
\frac{\partial f}{\partial t}+v \frac{\partial f}{\partial x}=
\frac{4\pi ^2 e^2 }{m^2}\frac{\partial}{\partial v}
\frac{W}{v}\frac{\partial f}{\partial v}, \label{eqk1}
\end{equation}
\begin{equation}
  \frac{\mbox{d}W}{\mbox{d}t}=\frac{\pi
\omega_{pe}}{n}v^2W\frac{\partial f}{\partial v}, \;\;\;
\omega_{pe}=kv \label{eqk2}
\end{equation}
 where $f(v,x,t)$ is the electron distribution function,
$W(v,x,t)$ is the spectral energy density of Langmuir waves.
$W(v,x,t)$ plays the same role for waves as the electron
distribution function does for particles. The system
(\ref{eqk1},\ref{eqk2}) describes the resonant interaction
$\omega_{pe}=kv$ of electrons and Langmuir waves, i.e. an electron
with the velocity $v$ can emit or absorb a Langmuir wave with the
phase velocity $v_{ph}=v$. In the right-hand side of equations
(\ref{eqk1},\ref{eqk2}) we omitted spontaneous terms due to their
smallness in comparison with induced effects \cite{Ryutov70}.

The evolution of the spectral function of the quasiparticles
$W(v,x,t)$ due to the spatial movement of Langmuir waves can be
described with the aid of the Liouville equation \cite{Camac62,
Vedenov67}
\begin{equation}\label{eqk2a}
  \frac{\mbox{d}W}{\mbox{d}t}=\frac{\partial W}{\partial t}+\frac{\partial
\omega}{\partial k} \frac {\partial W}{\partial x}-\frac{\partial
\omega_{pe}}{\partial x } \frac{\partial W}{\partial k},
\end{equation}
 where the two last terms in right hand side
of (\ref{eqk2a}) describe propagation of Langmuir waves and the
corresponding shift in wavenumber $k(x)$. Equation (\ref{eqk2a})
describes the propagation of Langmuir waves in the approximation
of geometrical optics or the WKB approximation \cite{Galeev63,
Vedenov67}. Clearly, this approximation imposes restrictions
\cite{Ryutov69}. The wavelength variation over the density
fluctuations scale should be a small fraction of itself
\begin{equation}\label{WKB}
\frac{\mbox{d}\lambda}{\mbox{d} x}\ll 1.
\end{equation}
Using the fact that the frequency of a Langmuir wave $\omega
(k(x), x)$ does not change over its propagation that implies a
wavenumber $k(x)$, which changes with position. Thus,
\begin{equation}\label{eq_o}
  \mbox{d}\omega (k(x), x)\equiv 0 = \left[\frac {\partial \omega(k,x)}{\partial
  x}\frac{\partial k}{\partial x}+\frac{\partial \omega _{pe}(k,x)}{\partial
  x}\right]\mbox{d}x,
\end{equation}
or
\begin{equation}\label{eq_k}
  \frac{\mbox{d} k(x)}{\mbox{d} x}= -\frac{\partial \omega }{\partial
  x}\left[\frac{\partial \omega}{\partial k}\right]^{-1}.
\end{equation}
Using $\lambda =1/k$, $k\approx \omega_{pe}/v$, $\omega (k,x)
=\omega _{pe}(x)[1+3k^2v_{Te}^2/2\omega_{pe}^2(x)]$ and eq.
(\ref{eq_k}) we have from equation (\ref{WKB})
\begin{equation}\label{cond}
  \frac{v}{|L|} \ll 3\omega_{pe}(x)\left(\frac{v_{Te}}{v}\right)^2
\end{equation}
where $L$ is the scale of the local inhomogeneity (\ref{drift}).
If the condition (\ref{cond}) is fulfilled than propagation of
Langmuir waves can be considered within the WKB approximation.

The clouds of fast electrons are formed in the spatially limited
regions of solar corona where acceleration takes place. Therefore,
spatially bounded beam is taken for consideration. The initial
electron distribution function is
\begin{equation}
f(v,x,t=0)=\frac{2n'}{\sqrt{\pi}\Delta v}
\mbox{exp}\left(-\frac{(v-v_0)^2}{\Delta v^2}
\right)\mbox{exp}(-x^2/d^2), \;\; v<v_0 \label{eq:4}
\end{equation}
where $d$ is the characteristic size of the electron cloud, $v_0$
is the velocity of the electron cloud, and $\Delta v\ll v_0$ is
the width of the beam in the velocity space. It is also implied
that initially the spectral energy density of Langmuir waves
\begin{equation}
W(v,x,t=0)=10^{-8}mn'v_0^3/\omega_{pe}, \label{eq:5}
\end{equation}
 is of the thermal level and uniformly distributed in space. The system
of kinetic equations (\ref{eqk1}-\ref{eqk2a}) is nonlinear with
three characteristic time scales. The first is the quasilinear
time $\tau (x)=n/n'\omega_{pe}$ that is determined by the
interaction of particles and waves. The second is the
characteristic time of Langmuir wave velocity shift due to plasma
gradient. The third scale is the time length of the electron cloud
$d/v_0 \gg \tau$. And we are interested in dynamics of the
electron cloud at the time scale $t \gg d/v_0$.

To solve kinetic equations of quasilinear theory we use finite
difference method \cite{Smith85,Samarskii89,Thomas95}. Each
differential operator is replaced by finite-difference one. To
achieve high accuracy of transport terms in space and velocity
space we use monotonic scheme \cite{vanLeer74, vanLeer77a,
vanLeer77b} that is considered to be the best in such kind of
problems \cite{Hawley1984}. To approximate the right hand terms we
use implicit finite difference scheme
\cite{Takakura82,Samarskii89}. This scheme seems to have better
results near plateau maximum velocity \cite{Kontar00}.

\section{Solar corona density model}

In order to obtain the density dependency of the solar corona we
follow the heliospheric density model \cite{Mann99}. On one hand
this is quite a simple isothermal model. However, it gives results
which agree with observations (see \cite{Mann99} for details).
These two features justify the choice of the model.

To derive the required heliospheric density model the
magnetohydrodynamic equations
\begin{equation}\label{hydro1}
\frac{\partial \rho}{\partial t}+ \mbox{div}(\rho \vec{v})=0
\end{equation}
\begin{equation}\label{hydro2}
\rho\left[\frac{\partial \vec{v}}{\partial t}+(\vec{v}
\nabla)\right]=-\nabla p-\frac{1}{4\pi}\left(\vec{B}\times
\mbox{rot}\vec{B}\right)+\vec{f}
\end{equation}
\begin{equation}\label{hydro3}
\frac{\partial \vec{B}}{\partial
t}=\mbox{rot}\left(\vec{v}\times\vec{B}\right)
\end{equation}
are employed with the velocity $\vec{v}$ of the flow, the mass
density $\rho$, the thermal pressure $p$, the magnetic field
$\vec{B}$, and the gravitational force $\vec{f}= \rho
GM_s\vec{n_r}/r^2$, where $G$- gravitational constant; $M_s$ - is
the mass of the Sun, $\vec{n_r}$ is the unit vector along the
radial direction. The system of equation
(\ref{hydro1}-\ref{hydro3}) is completed with the isothermal
equation of state
\begin{equation}\label{state}
  p = 1.92 n k_B T
\end{equation}
$n$ - electron number density, $k_B$ - Boltzmann's constant, $T$
is the temperature. The mass density $\rho$ and the electron
number density $n$ are related by $\rho = 1.92\tilde{\mu}m_pn$
($m_p$ is the proton mass). In the solar corona and the solar wind
$\tilde{\mu}$ has a value of $0.6$ \cite{Priest82}.

The stationary spherical symmetric solutions of the system
(\ref{hydro1}-\ref{state}) can be found. Assuming all the
variables to be radially directed equation (\ref{hydro1}) can be
integrated to
\begin{equation}\label{sol1}
r^2n(r)v(r)= C= const
\end{equation}
with the constant $C$. And equations (\ref{hydro2},\ref{hydro3})
can be transformed to the second equation
\begin{equation}\label{sol2}
  \frac{v(r)^2}{v_c^2}-\mbox{ln}\left(\frac{v(r)^2}{v_c^2}\right)=
  4\mbox{ln}\left(\frac{r}{r_c}\right)+4\frac{r_c}{r}-3
\end{equation}
with the critical velocity $v_c\equiv
v(r_c)=(k_BT/\tilde{\mu}m_p)^{1/2}$ and the critical radius
$r_c=GM_s/2v_c^2$, the distance where corona becomes supersonic.

\begin{figure}
\includegraphics[width=160mm]{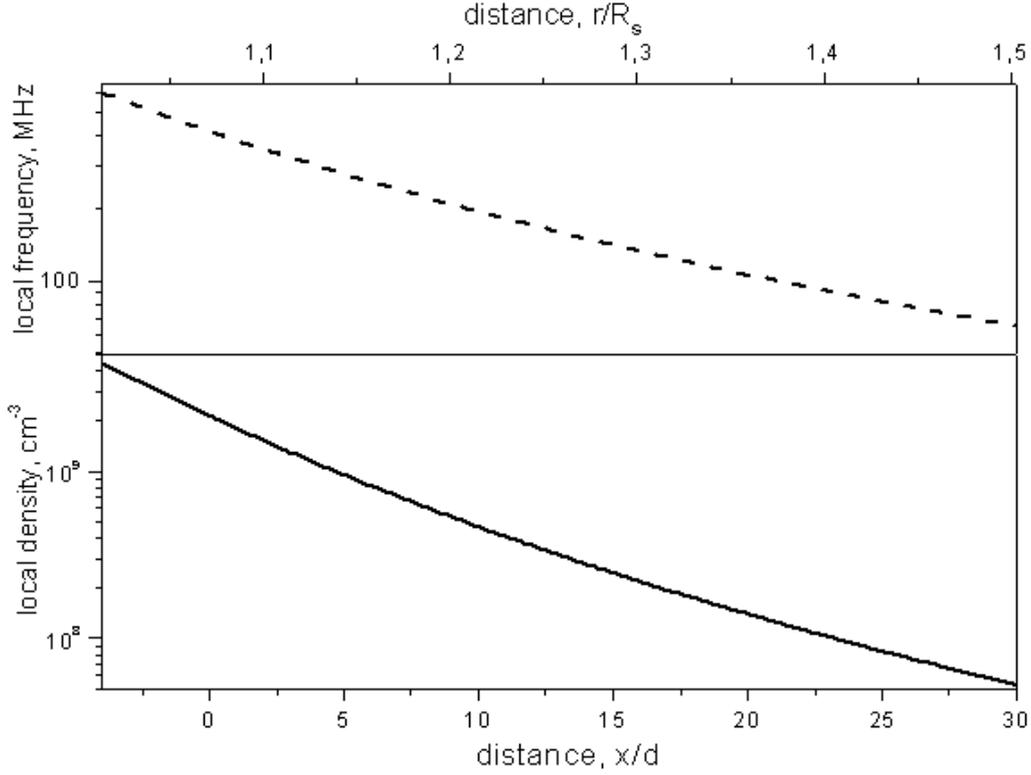}
 \caption{Density and local plasma frequency
$f_p(r)=\omega _{pe}/2\pi$ profiles. Numerical solution of
equations (\ref{sol1}, \ref{sol2}), where $r=R_s+x_0+x$ is the
distance from the center of the Sun, $x$ is the distance from the
initial position of the beam, and $R_s=6.958\times 10^{10}$cm is
the radius of the Sun. }
 \label{fig1}
\end{figure}

The density distribution $n(r)$ is found by numerical integration
of the equations (\ref{sol1},\ref{sol2}). The constant $C$ appearing
in (\ref{sol1}) is fixed by satellite measurements near the Earth
orbit (at $r=1$AU $n=6.59$~cm$^{-3}$ ). Thus following
Mann et al. (\citeyear{Mann99}), constant $C$ is determined
to be $C=6.3\times 10^{34}$s$^{-1}$.
For calculations we take the electron
temperature to be $10^6$K. To solve equations
(\ref{sol1},\ref{sol2}) numerically {\it rugula falsi} method is
used \cite{Korn61}. The density $n(r)$ and corresponding local
plasma frequency $f_p(r)$ profiles are presented in fig.
\ref{fig1}.

In the limit $v\ll v_c$, (\ref{sol1}) and (\ref{sol2}) can be
solved analytically in terms of a barometric height formula:
\begin{equation}\label{parker}
n(r) =
n_s\mbox{exp}\left[\frac{A}{R_s}\left(\frac{R_s}{r}-1\right)\right]
\end{equation}
with $A=\tilde{\mu}GM_s/k_BT$ and $n(r=R_s)=n_s=5.14\times
10^9$~cm$^{-3}$.

Note that the main factor ruling the density distribution at the
distances of our interest $r\leq 1.5R_s$ is the gravitational
force and the density distribution must be close to the barometric
formula. Therefore, the choice of the different model and
improvements of equations (\ref{hydro1}-\ref{state}) should not
lead to significant change in density profile.

\section{Results of numerical solution of kinetic equations}

The initial location of the source is taken to be $x_0=50000$km
\cite{Smith77}. In the paper we refer to the attitude $x_0$ as the
initial location of the electron cloud $x=0$. Electron beam is
assumed to be close to monoenergetic, $\Delta v=v_0/8$ and the
spatial size of the beam is taken to be $d=10^9$cm for all
calculations presented in the paper.

\subsection{Initial evolution of the electron beam}

The value determining the time of beam - plasma interaction and
consequently the character of beam propagation is the quasilinear
time. In the case of inhomogeneous plasma quasilinear time depends
on distance whereas in the uniform plasma it is a constant value.
When the quasilinear time is smaller than the time of electron
beam propagation we have quasilinear propagation and free
propagation in the opposite. If the initial electron beam density
so that $\tau (x=0)>t$ then electrons initially move without
interaction with plasma
\begin{equation}\label{eq_free}
f(v,x,t)=\frac{2n'}{\sqrt{\pi}\Delta v}\mbox{exp}
\left(-\frac{(v-v_0)^2}{\Delta v^2}\right)
\mbox{exp}(-(x-vt)^2/d^2), \;\;\; v<v_0
\end{equation}
As a result of electron propagation in plasma with decreasing
density quasilinear time
\begin{equation}\label{eq_qv}
  \tau (x,t) = \left(\omega
  _{pe}\int^{v_0}_{0}f(v,x,t)\mbox{d}v/n(x)\right)^{-1},
\end{equation}
becomes smaller. Thus, at the maximum of electron density
$x\approx v_0t$ equation (\ref{eq_qv}) becomes
\begin{equation}\label{eq_qv2}
  \tau (x,t) = \tau _0 \frac{d^2+(\Delta v/v)^2
  x^2}{d^2}\mbox{exp}\left(\frac{A}{2R_s}\left(\frac{R_s}{(R_s+x_0+x)-1}\right)\right)
\end{equation}
where $\tau _0=n(x=0)/(\omega_{p0}n')$, $\omega
_{p0}=\omega_{pe}(x=0)$, the plasma density and the local plasma
frequency are defined by barometric formula (\ref{parker}). Thus,
quasilinear relaxation appears for the first time when $\tau
(x,t)$ given by (\ref{eq_qv2}) becomes equal to $t$ - the time of
electron propagation.

\begin{figure}
\includegraphics[width=150mm]{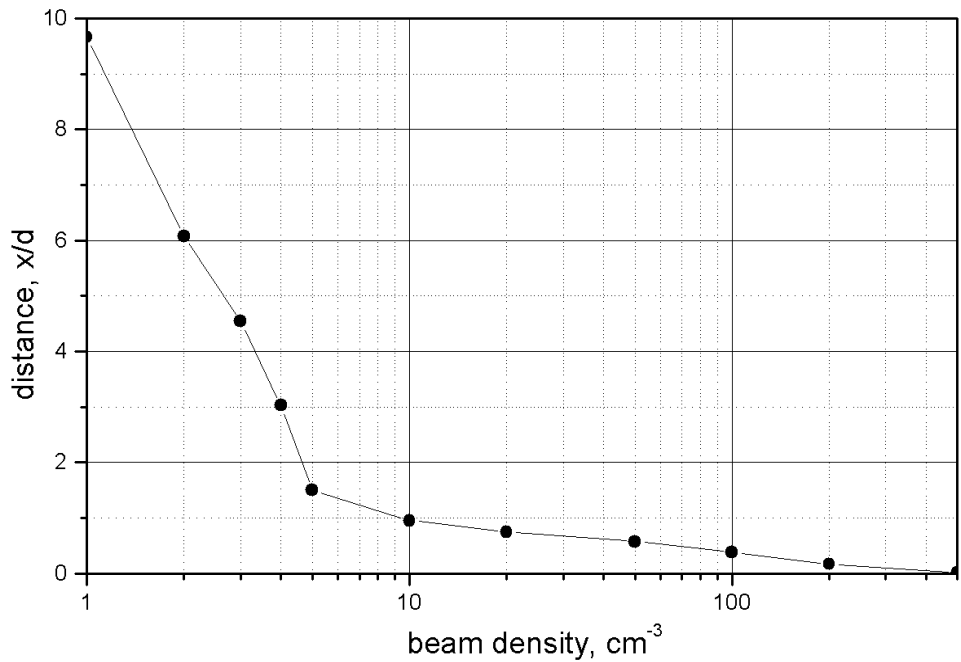}
 \caption{Starting frequency as the function of electron
 beam density. Numerical solution
 of kinetic equations $v_0=1\times10^{10}$cm~s$^{-1}$.}
 \label{fig9}
\end{figure}

Numerical solution of kinetic equations (\ref{eqk1}-\ref{eqk2a})
gives us that various electron beam densities lead to a different
points where relaxation occurs for the first time. In fig.
\ref{fig9} the
dependency on electron beam density
of the distance where Langmuir waves start to grow for
the first time and a plateau is formed in the electron distribution
function is presented. Later, we see
that the maximum of the Langmuir wave turbulence is located at
this point. We see the strong tendency of the first relaxation
point to decrease with density growth. Thus for beam densities
$n'\gtrsim 10$~cm$^{-3}$ relaxation and correspondingly the starting
frequency of bursts takes place near the initial electron beam location
$x\approx 0$. For beams of low density $n'<10$~cm$^{-3}$ the
distance between the initial location and relaxation point can be
 significant (about $10d$ for $n'=1$~cm$^{-3}$)
(fig.\ref{fig9}).

\subsection{The electron distribution function and the spectral
energy density of Langmuir waves}

\begin{figure}
\includegraphics[width=150mm]{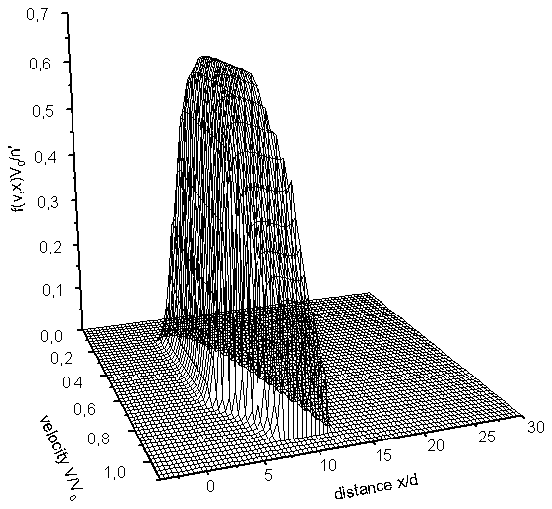}
\includegraphics[width=150mm]{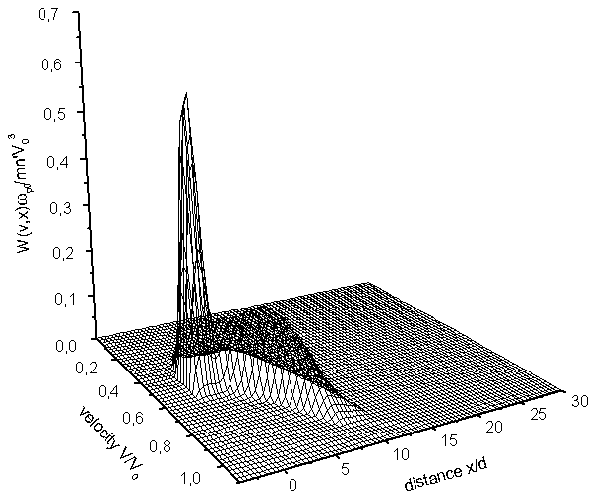}
 \caption{The electron distribution function and the spectral
 energy density of Langmuir waves at $t=1$s. Numerical
 solution of kinetic equations $n'=10$~cm$^{-3}$,
 $v_0=1.0\times 10^{10}$cm~s$^{-1}$.}
 \label{fig2}
\end{figure}

\begin{figure}
\includegraphics[width=150mm]{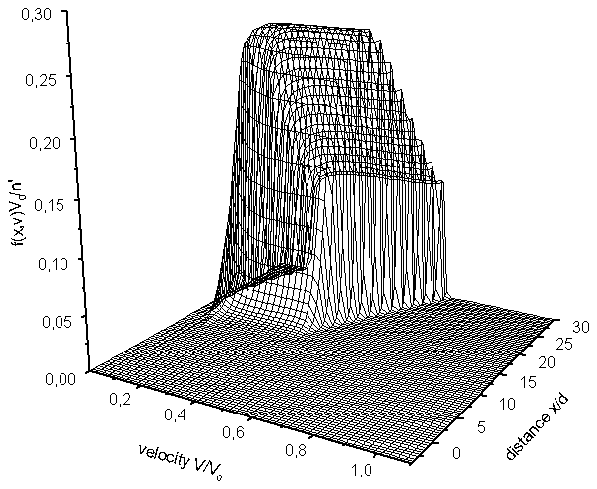}
\includegraphics[width=150mm]{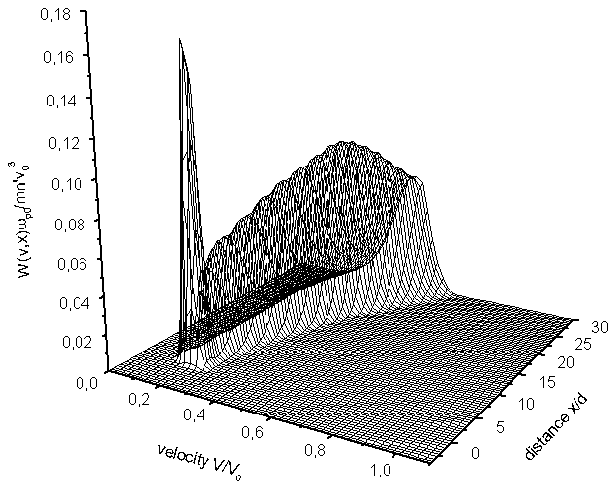}
 \caption{The electron distribution function and the spectral energy
 density of Langmuir waves at $t=5$s. Numerical solution of kinetic
 equations $n'=10$~cm$^{-3}$, $v_0=1.0\times 10^{10}$cm~s$^{-1}$.}
 \label{fig3}
\end{figure}

 The electron distribution function and the spectral energy density are
presented in fig. \ref{fig2},\ref{fig3}. In every spatial point
the electron distribution function is a plateau from almost zero
velocity up to some maximum velocity.  This maximum velocity is
not a constant value as it takes place in uniform plasma but a
decreasing function of distance. It is the direct result of plasma
inhomogeneity.
We also see that the slow spatial movement of Langmuir waves leads
to the shift of phase velocities toward smaller ones. In fig.
\ref{fig3} we see the result -- there is no plasma waves with the
phase velocity more than $0.7v_0$.

    The spectral energy density consists of two parts. The first
part is the Langmuir turbulence  located near initial beam
location $x\approx 0$ (see fig. \ref{fig2},\ref{fig3}). This part
of Langmuir waves drifts toward small phase velocities. Thus, the
peak is placed close to $v=0.5v_0$ at $t=1$s, while at $t=5$s the
peak is close to $v\approx 0.2v_0$ (initially the peak appears at
$v\approx v_0$ at $t\approx \tau (0)$). In the homogeneous plasma
the spectral energy density of Langmuir waves also consists of two
parts but the waves located near injection point are time
independent. Following Kontar, Lapshin and Melnik (\citeyear{Kontar98})
we have
\begin{eqnarray}\label{eq_peak}
  W_1(v,x) =\left\{\begin{array}{ll}
              \displaystyle \frac{m}{\omega _{pe}}\frac{n'}{v_0}v^5\mbox{exp}(-x^2/d^2),&\mbox{$v<v_0$}\\
              0,&\mbox{$v>v_0$}
              \end{array}
\right.
\end{eqnarray}
In the case of inhomogeneous plasma these waves (\ref{eq_peak})
evolve in accordance with equation (\ref{eqk2a}), where we can
neglect the spatial movement of plasma waves due to the smallness
of group velocity $\partial \omega/\partial k = 3v_{Te}^2/v\ll v$
and the interaction with electrons due to the electron absence in
this region
\begin{equation}\label{eq2}
  \frac{\partial W_1}{\partial t}-\frac{v^2}{L_0}
\frac{\partial W_1}{\partial v}=0,
\end{equation}
where $L_0=const=L(x=0)$. Integrating (\ref{eq2}) and using
(\ref{eq_peak}) as an initial condition we obtain the following
solution
\begin{eqnarray}\label{eq_peak2}
  W_1(v,x,t) \approx\left\{\begin{array}{ll}
              \displaystyle \frac{m}{\omega _{p0}}\frac{n'}{v_0}(1/v-t/L_0)^{-5}\mbox{exp}(-x^2/d^2),&\mbox{$v<u(t)$}\\
              0,&\mbox{$v>u(t)$}
              \end{array}
\right.
\end{eqnarray}
where $u(t)=v_0/(1+v_0t/L_0)$ is the maximum phase velocity for
these Langmuir waves. The equation (\ref{eq2}) is mathematically
correct and found in a good agreement with the numerical results.
Indeed, since the group velocity of Langmuir waves is a small but
a finite value $\partial \omega /\partial k<<v_0$, the spatial
movement of waves is hardly observable. Thus, at $t=1$s the peak
is close to $x\approx 0$, and only after $t=5$s the peak
approaches to $x\approx 2.5d$. However, the neglect of wave
movement violates the limitations of geometrical optics
(\ref{cond}), the velocity shift of Langmuir waves is only due to
the spatial wave movement.

The second part of Langmuir turbulence accompanies the electron
cloud. The electrons together with plasma waves present a
beam-plasma structure \cite{Melnik95, Kontar98}. These plasma
waves are also strongly influenced by the plasma density gradient.
Thus electrons arriving at a given point excite Langmuir waves. The
level of plasma waves is growing up to the moment when the maximum
of electron density arrives to this point. After this moment
electrons mainly with $\partial f/\partial v<0$ arrive to this
point that leads to absorption of Langmuir waves (see
Mel'nik (\citeyear{Melnik95}), Mel'nik, Lapshin and Kontar
(\citeyear{Melnik99}), Kontar, Lapshin and Mel'nik (\citeyear{Kontar98})
for details). Negligible
spatial movement of waves causes the plasma waves to move toward
smaller phase velocities. As a result the part of Langmuir waves
appeared to be out of resonance with particles. The back front has
more low velocity plasmons that can be absorbed by the electrons,
and the fastest electrons have no Langmuir waves with the
corresponding phase velocity to absorb. Therefore, a part of
beam-plasma structure energy is lost by the structure in the form
of low velocity Langmuir waves (compare fig. \ref{fig3}). In the
case of homogeneous plasma all the plasma waves generated at the
front are absorbed at the back of the structure
\cite{Melnik95,Kontar98}.

\subsection{Energy density distribution in electron and plasma wave subsystems}

\begin{figure}
\includegraphics[width=150mm]{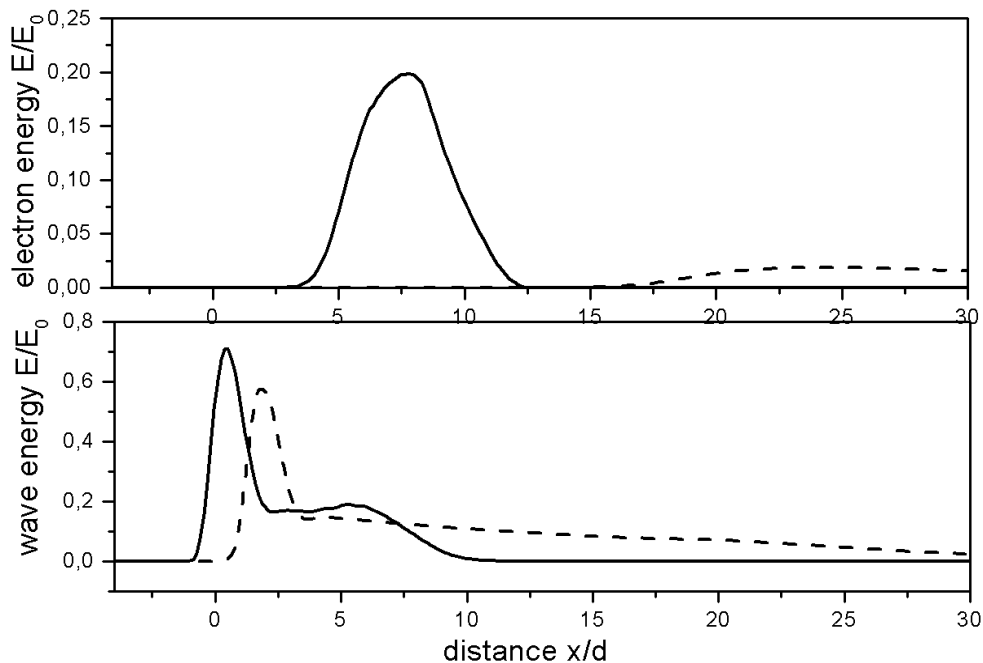}
 \caption{Langmuir wave and electron energy distribution at $t=1$s (solid lines)
 and at $t=5$s (dash lines).  Numerical solution
 of kinetic equations  $n'=10$~cm$^{-3}$, $v_0=1.0\times 10^{10}$cm~s$^{-1}$,
 $E_0= mn'v_0^2/4$ .}
 \label{fig6}
\end{figure}

The change of the spectrum of plasma waves influences the
energy distribution in the beam-plasma system. In the homogeneous
plasma beam-plasma structure propagates conserving its total
energy but if the density gradient is different from zero the
energy is pumped out of the structure. The balance between
electron and Langmuir wave subsystems is not observed in the
inhomogeneous plasma. The distribution of energy density is
presented in fig. \ref{fig6} at two different time moments.
The distribution of energy demonstrates that approximately a half
of the initial beam energy is concentrated near the site of
initial electron distribution $x\approx 0$ as it takes place in
the case of uniform plasma \cite{Melnik95,Kontar98,Melnik99}
\begin{equation}\label{eqe_0}
  E_1(x)/E(x) \equiv\int
  W(v,x,t)\mbox{d}k/E(x)=\omega_{pe}\int^{v_0}_{0}W_1(v,x,t)\mbox{d}v/E(x)=\frac{1}{2},
\end{equation}
where $E(x)= mn'v_0^2\mbox{exp}(-x^2/d^2)/2$ is the initial beam
energy density. The high level of Langmuir waves concentrated near
the initial beam location is due to initially nearly monoenergetic
electron beam (\ref{eq:4}). If the initial distribution function
(\ref{eq:4}) is proportional to $v$ \cite{Melnik99} then $W_1=0$
and the high level of Langmuir waves is not excited in the region of
initial relaxation.

The second part of the wave energy is the result of beam-plasma
structure propagation in the inhomogeneous plasma. The decrease in
the number of waves resonant with the beam causes the energy
losses. In fig. \ref{fig6} we see nonzero level of Langmuir waves
in the region where the structure has passed and the slow spatial
movement of plasma waves can also be seen. Despite of the
considerable energy losses the beam remains the source of a high
turbulence level (fig. \ref{fig6}).

The dissipation of energy by the beam-plasma structure changes the
spatial size of the beam-plasma plasma structure. The fast electrons,
which appeared to be out of resonance with waves can overlap the
structure electrons and pass some distance before they relax
toward a steady state (i.e. a plateau in the electron distribution
function). This effect increases the spatial size of the
structure.

\subsection{Velocity of a beam-plasma structure}

\begin{figure}
\includegraphics[width=150mm]{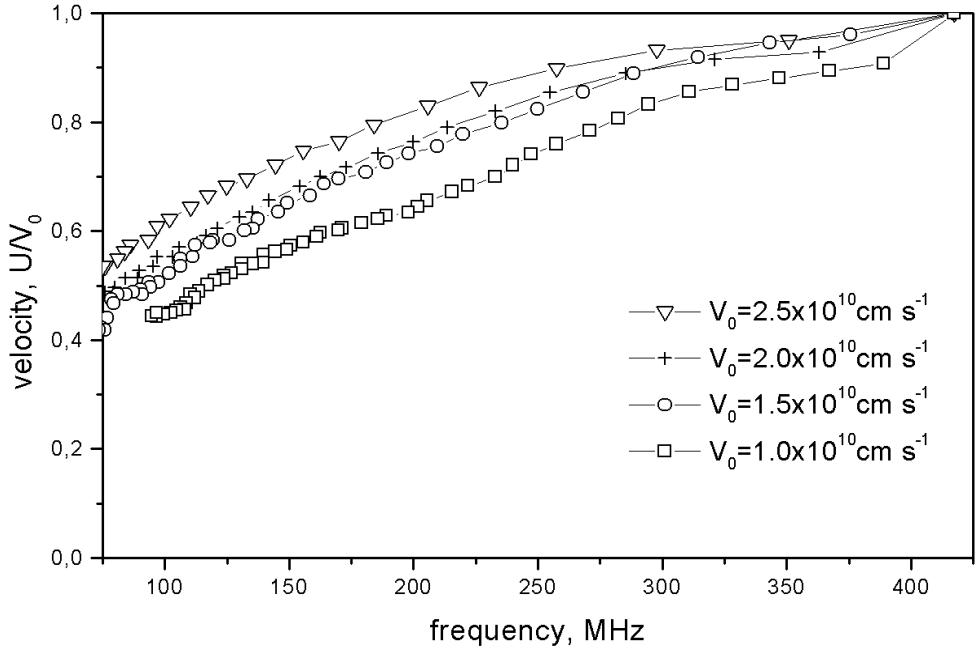}
 \caption{The dependency of maximum plateau velocity
 on local plasma frequency. Numerical solution
 of kinetic equations for $n'=10$~cm$^{-3}$ and various initial
 electron beam velocities.}
 \label{fig7}
\end{figure}

Recently it was shown that the beam-plasma structure propagates in
a uniform plasma with a constant velocity and the upper plateau
border (maximum plateau velocity) remains equal to the initial
value \cite{Melnik95,Kontar98}. The maximum velocity of the
plateau in the inhomogeneous plasma is not a constant. As we see
in the previous subsection the drift of the Langmuir waves leads to
the decrease of the maximum wave phase velocity. The electrons
with the velocity larger than the maximum wave velocity have no
resonance waves with the same phase velocity and therefore
propagate freely overtaking the interacting electrons. Thus the
fastest electrons leave behind the electrons of the plateau and
decrease the maximum velocity of the plateau i.e. decreasing the
average velocity of the electrons at a given point. The dependency
of the maximum plateau velocity measured at the electron density
maximum (this velocity corresponds to the velocity of a
beam-plasma structure) on distance is a decreasing function (fig.
\ref{fig7}). The speed of the structure can be approximated as
$v_{bps}\approx u/2$, where $u$ is the maximum velocity of the
plateau near the maximum of electron density. Thus we obtain that the
source of Langmuir waves propagates with a decreasing velocity.

\section{Discussion and main results}

From the physical point of view it is interesting to mention that
the main physical process leading to the change of the electron
distribution function and the spectrum of Langmuir waves is the
shift of the wavenumber $\Delta k$ (or phase velocity) due to
Langmuir wave spatial movement with group velocity. Despite  the
fact that the increment of beam-plasma instability depends on
distance it can be neglected in comparison with the first effect.
Indeed, as soon as we omit the terms responsible for the phase
velocity shift in kinetic equations the numerical solution becomes
very close to the homogeneous plasma case \cite{Kontar98,Melnik99}.
This numerical result is consistent with the qualitative conclusions
in \cite{Coste75}.

The results obtained in the previous section also bring us new
understanding for the theory of the radio bursts. Unfortunately we
do not directly observe the electron beams or Langmuir waves (the
exception is the satellite observations, which are far from the
initial beam acceleration site). However, following the standard
model \cite{Ginzburg58} that Langmuir waves are transformed into
observable radio emission via nonlinear plasma processes one
infers that the maximum of the spectral energy density of
Langmuir waves corresponds to the maximum of radio emission.

The numerical results show that electron beams with different
electron beam densities have initial relaxation point at various
distances. Electron beams with low density initially propagate
freely. However, the quasilinear time fastly decreases with distance
due to the decreasing plasma density. Therefore, at some distnace,
which is a function of beam density, quasilinear relaxation
appears. The
point where relaxation occurs for the first time give us the
starting frequency of the type III bursts. Normal type III bursts
usually start somewhere between 250 and 500 MHz \cite{Dulk85} and
we can see from fig. \ref{fig9} that specifically this range of
frequencies seems to be the most probable for typical beam
densities.

We also see that the electron beam propagates as a
beam-plasma structure, which is the source of Langmuir waves
and consequently radio emission. Observations give us that the
sources of type III bursts propagate with almost a constant
velocity, typically in the range $0.2c<v<0.6c$ \cite{Dulk85}.
Direct measurements of solar electrons by space probes  imply that
the velocity of the sources are slowing down \cite{Fainberg70}.
Radial variations of the exciter velocity are also required to
satisfy consistency with other observational parameters
\cite{Robinson92a}. The numerical results demonstrate that the
velocity of beam-plasma structure becomes less as the structure
propagates into a plasma whereas in the uniform plasma the speed
of the structure is a constant value \cite{Kontar98,Melnik99}.

\begin{figure}
\includegraphics[width=150mm]{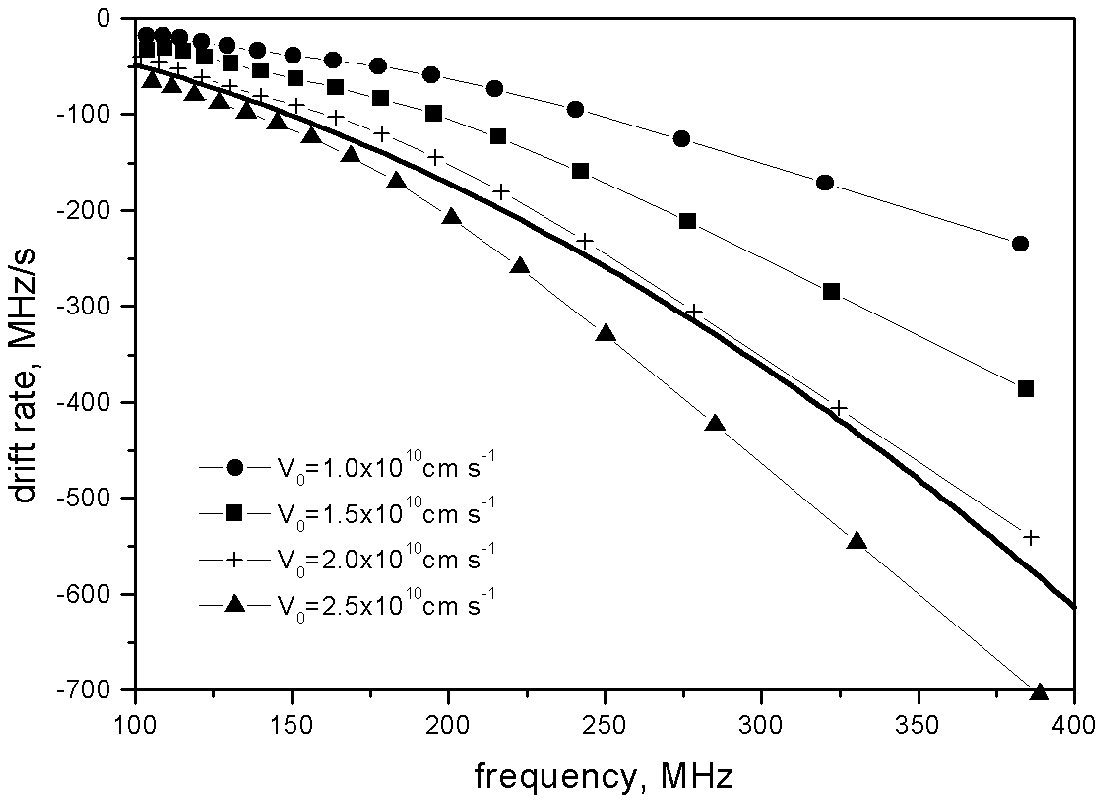}
\includegraphics[width=150mm]{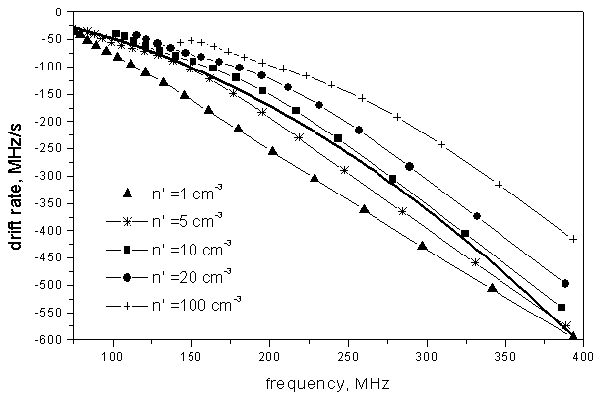}
 \caption{The frequency drift rate as a function of frequency.
 Upper figure - numerical solution
 for $n'=10$~cm$^{-3}$, and a various initial electron
 beam velocities; Lower figure - numerical solution for $v_0=2\times10^{10}$cm~s$^{-1}$,
 and for different beam densities. Solid line presents the observational
 fit (\ref{fit}).}
 \label{fig8}
\end{figure}

In observations the frequency drift rate is normally measured as a
function of frequency. Alvarez and Haddock (\citeyear{Alvarez73})
give
\begin{equation}\label{fit}
\frac{df}{dt}\approx -0.01f^{1.84}
\end{equation}
(where $f$ in Mhz, and $df/dt$ in MHz/s) the least-square fit to
the drift rates reported by various authors for the frequency
range from $550$MHz to $74$kHz. Indeed, assuming that the velocity
of the source is a constant and using (\ref{drift}) we cannot
explain the observational drift rate (\ref{fit}). To obtain the
observational properties of drift rate one should take into
account the deaceleration of electron beams. The plasma
inhomogeneity seems to be the physical process explaining the
drift rate observed in the range from 412 MHz to 70 MHz (fig.
\ref{fig8}). From fig. \ref{fig8} it could be concluded that the
''typical'' type III burst is generated by the electron beam with
the initial velocity $v_0=2\times10^{10}$cm~s$^{-1}$
(approximately $0.6c$) and density $n'=10$~cm$^{-3}$.

Our numerical results demonstrate that the main factor defining
the frequency drift rate is the initial electron density of the
beam (fig. \ref{fig8}). The initial velocity of the beam does play
a role but the impact on frequency dependency seems to be weaker
(see fig. \ref{fig8}).

Note, that as the velocity of the structure decreases with
structure propagation the decrease rate also declines. The highest
velocity decrease rate is near the starting  frequencies of type
III bursts, where the plasma gradient has the biggest value
(\ref{eq:gr}). Thus, the gradient of local plasma frequency
\begin{equation}\label{eq:gr}
  L(x)^{-1}=\frac{1}{\omega_{pe}}\frac{\partial \omega_{pe}(x)}{\partial x}=
- \frac{A}{2(R_s+x_0+x)^2}
\end{equation}
declines with the distance and respectively the influence of
plasma inhomogeneity becomes smaller. Therefore, one can expect
that the influence of plasma inhomogeneity on beam dynamics become
smaller further away from the initial beam location.

The second part of Langmuir turbulence generated by the electron
beam is located at the region of the initial electron beam
relaxation. Beam-plasma structure propagating in the inhomogeneous
plasma also leave behind quite high level of Langmuir turbulence.
These Langmuir waves propagate slowly in a plasma with
the group velocity $\sim 10^8$cm~s$^{-1}$. These waves can easy
transform into observable radio emission and might be responsible
for the type V bursts. Indeed, if we note that Langmuir waves
drift in velocity space due to inhomogeneity toward smaller phase
velocities than we might expect the decrease of the transformation
rate of Langmuir waves into observable radio emission and
consequently the long duration of the emission. Moreover, the
Langmuir waves drift very slowly and can be observed only in the range
from the starting frequencies of type III bursts to the distance
of a few solar radii, where plasma density gradient becomes much smaller.
Finally we should note that these plasma waves are determined
by the initial distribution function of the beam electrons.
The distribution, which is proportional to the velocity
\cite{Kontar98} gives no waves in the
initial beam location whereas monoenergetic beam leaves a half of
its kinetic energy in this site \cite{Melnik95}.

\section{Summary}

Dynamics of electron beam in the decreasing plasma density
demonstrate rather strong influence of plasma inhomogeneity. The
strongest influence is observed near starting frequencies of type
III solar radio bursts ($f_p>200$MHz). The beam of fast electrons
propagates in the corona plasma as a beam-plasma structure
\cite{Melnik95,Kontar98} that explains on one hand the ability of
beams to propagate over large distances being a source of plasma waves
and on the other hand the decrease of velocity of the burst
source. The plasma with declining density plays a role of
dissipative media, where beam-plasma structure in the course of
its propagation leaves a part of its energy in the form of
Langmuir waves. The comparison of the obtained drift rate with
observational data manifests a good agreement. The results
obtained allows us to quantitatively connect the some observable
properties of bursts (drift rate and starting frequency) with the
basic plasma an beam parameters (plasma density, density gradient,
beam density, and beam velocity).

However, more theoretical and numerical work is needed to enable
detailed comparison of the observational properties and
theoretical results. The brightness temperature, the size of the
source, radiation flux can not be determined without involving the
nonlinear plasma processes responsible for radio emission.
Moreover, solar corona plasma is not a monotonic function of
distance but rather inhomogeneous with different length scales.
All these and probably other processes should be included to
achieve quantitative understanding of electron beam dynamics.

\end{article}

\end{document}